\documentclass[preprint]{emulateapj}

\shorttitle{Spitzer observations of SN~2005af} 

\shortauthors{Kotak et al.}

\def\kms{\ifmmode{\rm km\,s^{-1}}\else\hbox{$\rm km\,s^{-1}$}\fi}

\begin{document}


\title{Spitzer measurements of atomic and molecular abundances in the Type IIP SN~2005af}


\author{
Rubina Kotak\altaffilmark{1},  
Peter Meikle\altaffilmark{2},
Monica Pozzo\altaffilmark{2},
Schuyler D. van Dyk\altaffilmark{3},
Duncan Farrah\altaffilmark{4},
Robert Fesen\altaffilmark{5},
Alexei V. Filippenko\altaffilmark{6},
Ryan Foley\altaffilmark{6},
Claes Fransson\altaffilmark{7},
Christopher L. Gerardy\altaffilmark{2},
Peter A. H\"{o}flich\altaffilmark{8},
Peter~Lundqvist\altaffilmark{7},
Seppo Mattila\altaffilmark{9},
Jesper Sollerman\altaffilmark{10}, \&
J.~Craig~Wheeler\altaffilmark{11}
}
\altaffiltext{1}{ESO, Karl-Schwarzschild Str. 2, Garching-bei-M\"{u}nchen, D-85748, 
                 Germany e-mail: {\tt rkotak@eso.org}}
\altaffiltext{2}{Astrophysics Group, Imperial College London, Blackett Laboratory,
               Prince Consort Road, London, SW7 2AZ, U.K.}
\altaffiltext{3}{Spitzer Science Center, 220-6, Pasadena, CA 91125}
\altaffiltext{4}{Department of Astronomy, Cornell University, 106 Space Sciences, Ithaca, 
                 NY 14853}
\altaffiltext{5}{Department of Physics and Astronomy, 6127 Wilder Lab., Dartmouth College, 
                 Hanover, NH 03755}
\altaffiltext{6}{Department of Astronomy, University of California, Berkeley, CA 94720-3411}
\altaffiltext{7}{Department of Astronomy, Stockholm University, AlbaNova, SE-10691 
                 Stockholm, Sweden}
\altaffiltext{8}{Department of Physics, Florida State University, 315 Keen Building,
                 Tallahassee, FL 32306-4350}
\altaffiltext{9}{Department of Physics and Astronomy, Queen's University Belfast, 
                 Belfast BT7 1NN}
\altaffiltext{10}{Dark Cosmology Centre, Niels Bohr Institute, University of Copenhagen, 
                 Juliane Maries Vej 30, 2100 Copenhagen {\O}, Denmark}
\altaffiltext{11}{Astronomy Department, University of Texas at Austin, Austin, TX 78712}




\begin{abstract}

We present results based on Spitzer Space Telescope mid-infrared 
(3.6-30\,$\mu$m) observations of the nearby IIP supernova 2005af. 
We report the first ever detection of the SiO molecule in a
Type IIP supernova. Together with the detection of the CO fundamental,
this is an exciting finding as it may signal the onset of dust 
condensation in the ejecta. From a wealth of fine-structure 
lines we provide abundance estimates for stable Ni, Ar, and Ne which,
via spectral synthesis, may be used to constrain nucleosynthesis models.


\end{abstract}


\section{Introduction}

Mid-infrared (mid-IR) studies of supernovae (SNe) have been hampered 
by a combination of several factors: their intrinsic faintness at epochs 
of interest, the high background at mid-IR wavelengths, and strong
terrestrial atmospheric absorption. With the sole exception of the
extremely nearby (50\,kpc) SN~1987A \citep{roche:93, wooden:93}, this 
wavelength region has remained out of reach for SN studies.
Yet, access to the mid-IR region is important for at least two
reasons. First, a large number of fine-structure lines that are
sensitive to the details of the explosion physics lie in the
5--40\,$\mu$m region.  These provide robust measurements of element
abundances permitting direct comparisons with explosion models.
Second, core-collapse SNe (CCSNe) have been proposed to be major
sources of dust at high redshifts \citep[e.g.,][]{tf:01}, but even
in the local Universe, this hypothesis rests on little direct 
observational evidence. By monitoring the mid-IR evolution, 
where warm dust emits most strongly, we have a powerful means
by which we can ascertain the epoch of dust condensation, and
estimate the amount of fresh dust produced in the ejecta.
Alternatively, via mid-IR measurements, we can study the
development of an IR echo arising in the pre-existing dusty
circumstellar medium, thereby constraining the pre-supernova 
evolution of the progenitor \citep[e.g.,][]{meikle:06}.

The improvements in terms of sensitivity and spatial resolution
afforded by the Spitzer Space Telescope ({\it Spitzer}) finally allow
us to probe SN behaviour using mid-IR diagnostics. To this end, we
are pursuing a multi-epoch spectroscopic and photometric campaign of 
a significant sample of SNe of all types, within the framework of the
Mid-IR Supernova Consortium (MISC). Fortuitously, since the launch of
{\it Spitzer}, at least half a dozen CCSNe have exploded within 
$\sim$10\,Mpc.
Recently, we presented a first mid-IR study of the nearest (3.13\,Mpc)
of these, SN~2004dj \citep{kotak:05}.  Here, we describe results for
the Type IIP SN~2005af that highlight the diversity of this
subgroup.

SN~2005af was discovered in the nearby edge-on spiral galaxy, NGC~4945, on 2005 Feb. 
8.22 (UT) at V$=+12.8$ by \citet{jp:05}. On the basis of a spectrum obtained four 
days later, \citet{ff:05} classified it as a Type~IIP (plateau) supernova, at an 
epoch of about one month post-explosion. In what follows, we assume that the supernova 
exploded on 2005 Jan. 9, but note that the explosion epoch is probably uncertain 
by up to a few weeks. NGC~4945 is associated with the Centaurus group of galaxies, 
and estimates of its distance have ranged from 3.9 \citep{dv:81} to 8.1\,Mpc \citep{baan:85}.
In recent years, these values have tended to converge near the lower end of this range. 
Following the reasoning in \citet[][and references therein]{bergman:92}, we adopt a 
distance of 3.9\,Mpc.

\section{Observations}

We first observed SN~2005af with {\it Spitzer} during Cycle~1 using
the Infrared Spectrograph (IRS) and IRS Peak-up Imaging (PUI) under
Director's Discretionary Time Program 237. We subsequently observed 
it during Cycle~2 as part of our GO Program 20256, using the Infrared 
Array Camera (IRAC), IRS, and IRS-PUI.

\subsection{Photometry}

3.6--8.0\,$\mu$m imaging with IRAC was obtained on 2005 Jul. 22 (194\,d)
and 2006 Feb. 12 (399\,d). 
A frame time of 30\,s  and 5 (medium-sized) dither positions in mapping mode 
for each pair of channels were used. Our measurements from aperture photometry, 
carried out on the post-Basic Calibrated Data, vS14.0.0 image 
mosaics using GAIA, are shown in Table \ref{tab:phot}.
At 67\,d, SN~2005af is very bright in all four IRAC channels, The
flux is particularly bright in channel~2 (4.5\,$\mu$m).  Nevertheless,
examination of individual BCD frames showed that saturation had
not occurred i.e. the measurements are robust.

Our first epoch of PUI-16\,$\mu$m imaging was obtained in Peak-Up mode
only, i.e., before dithering capabilities were offered with IRS. 
The two subsequent epochs of PUI observations were
carried out using 31.46\,s ramps and 20 small-scale dither positions.
The SN was clearly detected at all three epochs (Table \ref{tab:phot}).

\begin{figure}
\begin{centering}
\includegraphics[width=0.47\textwidth,clip=]{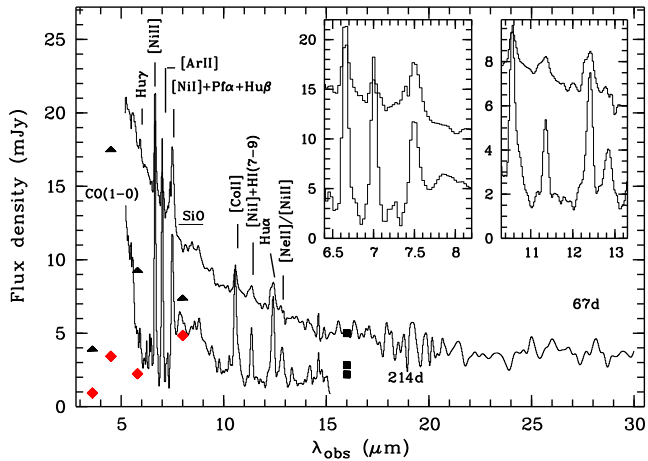}
\end{centering}
\caption{Mid-IR spectra of SN~2005af for days 67 ({\it top curve}) and 214 
         ({\it bottom curve}). Plausible line identifications are marked. 
         The IRAC measurements are shown as triangles (194\,d) and
         ({\it red}) diamonds (399\,d), while the black squares show 
         the PUI measurements for 67, 211, 433\,d from top to bottom, respectively.
         Note the sharp drop in flux between the two IRAC epochs at 4.5\,$\mu$m, 
         and the relative increase in prominence of the flux at $\sim$8\,$\mu$m.  
         The IRS flux levels are consistent with the photometry. The insets show 
         the evolution of line profiles described in $\S$\ref{sec:analysis}.
         \label{fig:irsboth}
        }
\end{figure}

\subsection{Spectroscopy}

Spectra were obtained on 2005 Mar. 17 (67\,d) and 2005 Aug. 11
(214\,d) with the Short-Low (SL, $\lambda = 5.2-14.5\,\mu$m; R = 64-128)
module of the IRS in staring mode. A Long-Low (LL, $\lambda =
14-38\,\mu$m; R = 64-128) spectrum was also acquired at 67\,d.  For the
SL mode, the total time-on-source at the two epochs was 1219\,s and
1828.5\,s respectively, and 629\,s for the LL mode. 

%
The first and second epoch data were preprocessed using versions
S12.3.0 and S12.4.0, respectively, of the {\it Spitzer} data processing
pipeline. We began the reduction of the BCD data by differencing the
two nod positions to remove the background emission. All subsequent
steps, i.e., straightening, extraction, wavelength and flux calibration
were carried out using SPICE v1.1-beta15 with standard settings
The spectra are shown in Fig. \ref{fig:irsboth}. We also 
extracted the spectra using post-BCD S13.2.0 products and found 
that they were virtually identical.

\section{Analysis}

\label{sec:analysis}
The high IRAC flux at 4.5\,$\mu$m for the first epoch compared to other
channels is almost certainly due to CO fundamental band emission.
This is corroborated by the steep rise towards 5\,$\mu$m in both
spectra, which we identify with the red wing of the CO fundamental
(see Fig. \ref{fig:irsboth}). Between 194\,d and 399\,d, the
sharpest decline in IRAC flux is seen at 4.5\,$\mu$m. These observations
suggest that CO formed in the ejecta less than 2 months after
explosion, persisting to at least 7 months, but had faded
substantially by about 1~year. 
The relative increase in the IRAC channel~4 (8\,$\mu$m) flux by
399\,d is probably due to the emergence of spectral features in
the 6.43--9.40~$\mu$m region spanned by this channel.  This is
discussed below.

By day 214, a pedestal-like feature appears spanning
$\sim$7.6-9\,$\mu$m, which we identify with molecular emission due to
the SiO fundamental band (Figs. \ref{fig:irsboth},
\ref{fig:compare}). This is the first time that SiO has been detected
in the ejecta of a SN~IIP -- the most common subtype of all
CCSNe at low redshifts. The only other detection of SiO in SN ejecta
was in the peculiar Type II SN~1987A \citep{roche:91,wooden:93}. That 
molecules are detected at all, provides a diagnostic of the prevalent 
conditions in the ejecta. As both CO and SiO are powerful coolants, 
they are regarded as precursors to dust condensation in the ejecta
\citep[e.g.][]{nozawa:03}.

\begin{figure}
\begin{centering}
\includegraphics[height=0.31\textwidth,clip=]{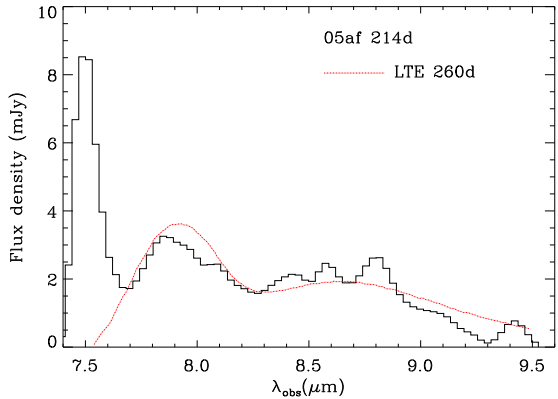}
\end{centering}
\caption{The 260\,d LTE model from \citet{ld:94}, attenuated by a factor of
         2.5, superimposed on the continuum-subtracted second-epoch spectrum of 
         SN~2005af. From this, we deduce a SiO mass of $\sim 2\times 10^{-4}$
         $M_\odot$ (\S\ref{sec:analysis}). }
\label{fig:model}
\end{figure}

\begin{figure}
\begin{centering}
\includegraphics[width=0.45\textwidth,clip=]{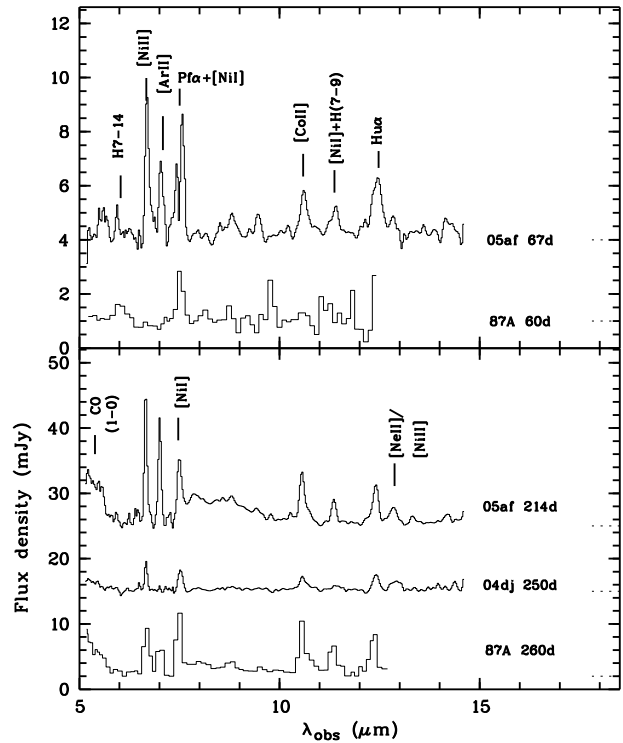}
\end{centering}
\caption{Comparison of continuum-subtracted spectra of CCSNe at 
         roughly coeval epochs. The dotted lines mark the zero-level.  
         Note the striking difference in the strength of the [ArII] line 
         at $\sim$200\,d between the SNe. The SN~1987A
         spectra are from \citet{wooden:93}; the 260\,d spectrum has
         been divided by a factor of 5. Isolated features that are common
         to both epochs are labelled in the top panel only; features that
         are evident in the later epoch only are indicated in the lower
         panel. 
         The SN~2004dj spectrum is part of our ongoing
         monitoring programme; a fuller analysis of these data will be
         presented elsewhere.  
         \label{fig:compare}}
\end{figure}

In Fig. \ref{fig:model} we compare an LTE model SiO spectrum 
for SN~1987A at 260~days \citep{ld:94} with the day~214 SL spectrum of
SN~2005af. The model, attenuated by a factor of 2.5, produces a convincing
reproduction of the observed behaviour. Scaling from the 
model, the match to our spectrum suggests a SiO mass of
$\sim$2$\times10^{-4}$\,M$_{\odot}$. \citet{ld:94,ld:96} argue that
such a mass can only form if there is no mixing of helium into the
oxygen-silicon region of the ejecta. 

Apart from molecular emission, several conspicuous lines are visible 
in both spectra (Fig. \ref{fig:irsboth}) which were identified 
by analogy with SN~1987A \citep{wooden:93}. At both epochs, the 
spectrum is dominated by forbidden lines of Ni, Co, and 
Ar, of which the [NiII] 6.64\,$\mu$m, [ArII] 6.99\,$\mu$m, 
and [CoII] 10.52\,$\mu$m are the most prominent. 

Between 67\,d and 214\,d the lines show appreciable increases 
in intensity (see Table \ref{tab:intensities}, Fig. \ref{fig:irsboth}),
In addition, in spite of the low spectral resolution (R=64-128), we
observe in a number of lines (e.g., [NiII] $\lambda$6.64\,$\mu$m,
[CoII]$\lambda$ 10.52\,$\mu$m) an extended red wing at 67\,d which 
fades by 214\,d. The broad blend with an extended red wing near
11.3\,$\mu$m at 67\,d is probably due to a combination of [NiI] and 
HI (7-9), but by 214\,d, it is symmetric.
Extended red wings were also detected in several lines in SN~1987A
\citep[e.g.,][]{witteborn:89} accompanied by small redshifts of the
line centroid. A plausible explanation is provided by electron
scattering in the expanding hydrogen envelope \citep{fc:89}.  This
may also apply to SN~2005af, although we do not detect any obvious
redshifting of line centroids.


We now estimate the abundances from the stronger lines at the later epoch 
when the ejecta has a lower optical depth. 
Following \citet{mccray:90}, we estimate the electron density of the 
ejecta as $n_e\sim10^{8}t_{\mathrm{yr}}^{-3.5}$ cm$^{-3} \approx 7\times10^{8}$
cm$^{-3}$ at day 214. 
We derived critical densities for Ni$^{+}$ and Ar$^{+}$ using collision 
strengths from \citet{bp:96} and \citet{osterbrock:06}, respectively,
and found these to be 1-3 orders of magnitude smaller than $n_e$.
On this basis, we deem an LTE treatment to be valid. 
%
Using the line intensities in Table \ref{tab:intensities}, we derive
masses using the upper excited state populations of the ions and find
3.7$\times 10^{-3}\,M_\odot$, 1$\times 10^{-3}\,M_\odot$, and
2.2$\times 10^{-3}\,M_\odot$ for stable (Ni$^0$,Ni$^+$), Ar$^+$, and
Ne$^+$ respectively, for an electron temperature of 4000\,K. Note that the 
result is insensitive to temperature: a $\pm$1000\,K change in temperature, 
changes the mass estimates by $\lesssim$15\%. The partition functions are from 
\citet{halenka:01} for Ni, and \citet{irwin:81} for Ar and Ne. 
Assuming a uniform distribution of stable Ni, we find that the Sobolev
optical depths of all detected Ni lines are less than 0.06. Strong transitions 
of [NiIII] 11.002\,$\mu$m and [NiIV] 8.405\,$\mu$m lie within the wavelength 
region covered, but these lines are not detected at either epoch. We
conclude that the Ni is predominantly in the neutral and singly-ionised 
state. Thus, assuming insignificant clumping, the value we derive above is 
probably close to the total stable Ni mass. 

The predicted mass of stable Ni is sensitive to the progenitor mass, 
although this relation is not monotonic, and depends sensitively on
the explosion mechanism, multi-dimensional effects and, in spherical
models, on the mass cut, which is a free parameter.  \citet{thielemann:96} 
predict stable Ni masses in the range 0.01--0.014$\,M_\odot$ for progenitors 
in the range 13-20$\,M_\odot$, and a much lower value of 0.002$\,M_\odot$ 
for a 25$\,M_\odot$ star. Other authors \citep[e.g.][]{cl:04,ww:95,nomoto:06}
derive comparable values, though detailed intercomparisons are difficult
due to different parametrizations of physical processes. 

\citet{cl:04} provide explosive yields as a function of metallicity
based on parameters which have been calibrated to fit observed properties of 
massive stars. The value for the stable Ni mass that we derive is closest to 
their progenitor models of 13-15\,$M_\odot$ with metallicities of a third to 
a twentieth of the solar value. Although we can definitively rule out higher 
mass progenitors at even lower metallicities, we current cannot exclude a 
35\,$M_\odot$ progenitor at solar metallicity. Further, planned, late-time
spectroscopy should provide tighter constraints.

Given the high energy required to ionise neutral Ne (21.6eV) and Ar
(15.8eV) we conclude that, given the low ionisation conditions
indicated by Ni, the bulk of the Ne and Ar lies in the
neutral state. However, no strong transitions of [NeI] or [ArI] lie in
our wavelength range. Consequently, the masses derived here for the
first ionisation state represent lower limits only.


For SN~1987A at 260~days, using the [CoII] 10.52\,$\mu$m
line, \citet{roche:93} derive a value for $M_{\mathrm{Co}^{+}}$ of
$4.6\times10^{-3}\,M_\odot$.  Using our day 214 intensity for this
line in SN~2005af we obtain $M_{\mathrm{Co}^{+}}$ =
$2.4\times10^{-3}\,M_\odot$. Allowing for the different epochs for
the two SNe, this implies that a $^{56}$Ni mass of 0.027\,$M_\odot$
was ejected by SN~2005af. This estimate falls in the expected range 
for normal Type~II SNe \citep[e.g.,][]{hamuy:03}.


Although we have a clear detection of the SN~2005af spectrum between
16-30\,$\mu$m (Fig.  \ref{fig:irsboth}) -- the first, for any SN at this
early epoch -- we do not see any significant spectral features. This
is not surprising, given its youth. In SN~1987A, lines of [FeII] in
this spectral region were only detected after $\sim$400\,d
\citep{haas:90}.

\subsection{Comparison with other Type~II SNe}
\label{sec:compare}

In Fig. \ref{fig:compare} we compare roughly coeval mid-IR spectra of
SN~2005af with that of SN~1987A (IIpec) at $\sim$60\,d, and with those
of SN~1987A and SN~2004dj (IIP) at $\sim$250\,d. The 60\,d
mid-IR spectrum of SN~1987A is dominated by HI lines
\citep[see Fig. \ref{fig:compare} or][]{roche:93,wooden:93}. In contrast, 
strong lines of [NiII] and [ArII] are present in SN~2005af. This suggests
either that our first-epoch SN~2005af observations were taken on the
radioactive tail of the lightcurve and that the length of the plateau
phase must have been $\lesssim$60\,d, or that the estimated explosion
date, based on the classification spectrum is too late by $\ga$1 month.

At the later epoch all three SNe exhibit prominent lines of nickel and
cobalt. \citet{wooden:93} emphasize that the ratio of [NiI]\,7.5\,$\mu$m
to [NiII]\,6.64\,$\mu$m lines provides a sensitive measure of the ionization
fraction of Ni ($\chi_{\mathrm{Ni}}=n_{\mathrm{Ni}^{+}}/n_\mathrm{Ni}$). 
Using their Fig. 9, and assuming LTE, we find $\chi_{\mathrm{Ni}} 
\approx 0.35, 0.55,$ and 0.85 for SN~2004dj, SN~2005af, and SN~1987A, 
respectively, at $\sim$250\,d.

SN~2005af shows much stronger [ArII] than SN~1987A. 
Moreover, SN~2004dj is found to be a factor of 20 weaker than SN~2005af 
in this line. Even if we normalise to the strengths of the adjacent 
[NiII]6.63\,$\mu$m, [NiI]7.50\,$\mu$m lines, the [ArII] line is still 
over $\times$5 stronger in SN~2005af. This is intriguing given that
SNe~2004dj and 2005af are both of Type~IIP. 
One possibility, indirectly implied by $\chi_{\mathrm{Ni}}$ above, 
is that the conditions required to ionize Ar are not attained
in SN~2004dj, while Ni$^+$, with its lower ionization potential (7.6eV),
is easier to produce.

Another hint may come from the spherically symmetric \citet{thielemann:96} 
models, where the bulk of the stable Ar (mostly $^{36}$Ar) is formed in 
a shell just above the stable Ni region. Thus, one might expect that a higher 
mass cut (below which the core collapses to a neutron star or black hole) 
within the Ni zone would produce a lower absolute Ni mass, but a higher Ar/Ni 
mass ratio. However, this seems to be contrary to what we observe, i.e.,
SN~2004dj has an apparently smaller Ni mass but an even smaller
Ar/Ni mass ratio than SN~2005af. We suspect that to account for this
behaviour, a certain degree of asymmetry or clumpiness, as well as mixing
in the ejecta has to be invoked. Interestingly, \citet{pereyra:06}
infer an overall asphericity of 20\% for SN~2005af from optical polarimetry.

\vspace*{-0.7cm}
\acknowledgments
\noindent
This work is based on observations made with the {\it Spitzer Space Telescope\/}, 
which is operated by the Jet Propulsion Laboratory, California Institute of 
Technology under NASA contract 1407. Financial support for the research
is provided by NASA/{\it Spitzer} grant GO-20256.
Support for this work was provided by NASA through an award issued by JPL/Caltech. 
R.K. and S.M. acknowledge support from an ESO fellowship and from the EURYI 
Awards scheme respectively.

\clearpage

\begin{table}
\caption{Photometry of SN~2005af}
\begin{center}
\begin{tabular}{ccclccccccclc}
\hline \tableline
        &  & & \multicolumn{5}{c}{Flux (mJy)} \\
\cline{4-8}
        &  Epoch & t$_{\mathrm{exp}}$ & \multicolumn{4}{c}{IRAC} &   \multicolumn{1}{c}{PUI} \\
Date    & (d)    & (s) & 3.6\,$\mu$m  & 4.5$\mu$m & 5.8\,$\mu$m  & 8.0\,$\mu$m &  16\,$\mu$m \\
\tableline
2005 Mar. 17 & \phn67 & 1219  & \nodata & \nodata & \nodata & \nodata & 5.0$\pm$0.3 \\
2005 Jul. 22 &   194  & \phn524  & 3.96$\pm$0.02 & 17.54$\pm$0.04 & 9.31$\pm$0.05 & 7.43$\pm$0.06  & [3.10] \\
2005 Aug. 08 &   211  & \phn629  & \nodata & \nodata & \nodata & \nodata & 2.83$\pm$0.03 \\
2006 Feb. 12 &   399  & \phn524  & 0.93$\pm$0.02 & 3.43$\pm$0.02 & 2.24$\pm$0.02 & 4.84$\pm$0.07   & [2.63]  \\
2006 Mar. 18 &   433  & \phn629  & \nodata & \nodata & \nodata & \nodata & 2.19$\pm$0.03 \\
\tableline
\end{tabular}
\end{center}
\tablecomments{
Photometry was carried out using a 5\farcs5 radius aperture for all channels.
This aperture was chosen to enclose the full diffraction pattern to the first dark
ring at the PUI red limit of 18.7\,$\mu$m. The aperture was positioned using WCS
coordinates. 
Aperture corrections as appropriate for the outer diffraction rings were applied 
to IRAC channels 1-4 using Table 5.7 of the IRAC data handbook. 
No attempt was made to apply a colour correction to the measured PUI fluxes, since 
the choice of temperature is not obvious. However, we note that the correction
would be $\lesssim$2\% for hot dust (500-1000\,K), and that such a correction would 
lie well within our errors. We estimated the systematic error due to variations in the 
background by repeating our measurements with differing annuli. The errors above are 
a combination of systematic and statistical errors. The figures in square brackets 
under the PUI column are indications of the expected fluxes coeval with the IRAC 
values assuming a $Q$-band decline rate similar to that of SN~1987A.
t$_{\mathrm{exp}}$ is the on-source integration time. 
\label{tab:phot}}
\end{table}


\begin{table}
\begin{center}
\caption{Line intensities \label{tab:intensities}}
\begin{tabular}{clccccccccc}
\tableline\tableline
            &          &               &             \multicolumn{2}{c}{67\,d}   & \multicolumn{2}{c}{214\,d} &   \\
\cline{4-5} \cline{6-7}
$\lambda_{\mathrm{rest}}$   &   Ion    &     Inst.Res. &  I	& FWHM    &   I  & FWHM   \\ 
($\mu$m)    &          &     (\kms)    &              & (\kms)  &      & (\kms) \\
\tableline
                    &         &      &                &          &         &       \\
6.636               & [NiII]  & 2700 &    3.18        &  4000    &  9.42   &  3600 \\
6.985               & [ArII]  & 2600 &    1.61        &  3800    &  6.16   &  3000 \\
7.507               & [NiI]   & 2400 &   \nodata      & \nodata  &  4.48   &  4000 \\
10.521$^\dag$       & [CoII]  & 3400 &    0.78        &  4500    &  1.99   &  3000 \\
11.308$^\ddag$      & [NiI]   & 3200 &    0.25        &  2400    &  0.78   &  2900 \\
12.813$^\star$      & [NeII]  & 2800 &    \nodata     &  \nodata &  0.56   &  3300 \\
\tableline
\end{tabular}
\end{center}
\tablecomments{
        The intensities were measured after subtracting a continuum
        fitted by a low-order polynomial or a spline function; the
        FWHM were derived from Gaussian fits. Column 3 lists the 
        instrumental resolution; I, the intensity, is in units of
        $10^{-14}$\,erg\,cm$^{-2}$\,s$^{-1}$. $^\dag$ As this line is
        symmetric, we assumed that there was no blend with [NiII]
        10.68\,$\mu$m. $^\ddag$ At 67\,d this line has a probable
        contribution from HI(7-9) but by day 214, it is symmetric;
        above, we assume that it arises due to [NiI] only. Interestingly,
        the HI(7-9) feature was not detected in SN~1987A at 60\,d 
        \citep{wooden:93}.
        $^\star$ The 12.8\,$\mu$m feature at 214\,d is likely to be 
        a blend of [NeII]\,$\lambda$12.813\,$\mu$m and [NiII]\,$\lambda$
        12.729\,$\mu$m. Given the strength of the [NiII] 6.636\,$\mu$m
        line, we estimate that the 12.8\,$\mu$m blend comprises roughly 
        equal contributions from the two elements. At 67\,d, however,
        given the strength of the [NiII] 6.636\,$\mu$m line, the [NiII]
        12.729\,$\mu$m line would have been barely detectable.
        }
\end{table}




\end{document}